\documentstyle[iopconf1,axodraw]{article}

\newcommand{\beq}{\begin{equation}}
\newcommand{\eeq}{\end{equation}}
\newcommand{\beqa}{\begin{eqnarray}}
\newcommand{\eeqa}{\end{eqnarray}}
\newcommand{\snu}{\tilde \nu}
\newcommand{\msnu}{m_{\tilde \nu}}

\newcommand{\dmsnu}{{\mbox{$\Delta m_{\tilde \nu}$}}}

\newcommand{\BR}{\mbox{BR}}

\def\gtrsim{{~\raise.15em\hbox{$>$}\kern-.85em\lower.35em\hbox{$\sim$}~}}
\def\lesssim{{~\raise.15em\hbox{$<$}\kern-.85em\lower.35em\hbox{$\sim$}~}}

\def\npb#1{{\sl Nucl.\ Phys.}\ {\bf B#1}}
\def\plb#1{{\sl Phys.\ Lett.}\ {\bf B#1}}
\def\prd#1{{\sl Phys.\ Rev.}\ {\bf D#1}}
\def\prl#1{{\sl Phys.\ Rev.\ Lett.} {\bf #1}}

\def\prep#1{{\sl Phys.\ Rep.}\ {\bf #1}}

\def\ifmath#1{\relax\ifmmode #1\else $#1$\fi}
\def\half{\ifmath{{\textstyle{1 \over 2}}}}

\begin{document}

\hfill\hbox{\vbox{
      \hbox{SLAC-PUB-7671}
      \hbox{hep-ph/9710276}}}

\def\myrule{\medskip\hrule height 1pt\bigskip}  

\renewcommand{\thefootnote}{\fnsymbol{footnote}}

\title{Sneutrino Mixing}
\footnotetext{ \hspace{-20pt}
\hrulefill \hspace{230pt} \par 
Talk presented 
at ``Beyond the Desert'' Workshop, Castle Ringberg, 
Tegernsee, Germany, 8-14 June 1997.}
\author{Yuval Grossman}
Stanford Linear Accelerator Center \\
        Stanford University, Stanford, CA 94309

\beginabstract
In supersymmetric models with nonvanishing 
Majorana neutrino masses, the sneutrino and antisneutrino mix.  
The conditions under which this mixing is
experimentally observable are studied, and 
mass-splitting of the sneutrino mass eigenstates and sneutrino
oscillation phenomena are analyzed.
\endabstract


\section{Introduction}
In the minimal Standard Model, as well as
in the minimal supersymmetric extension of it (MSSM)
neutrinos are exactly massless \cite{revnu}. 
The present direct laboratory upper bounds on their masses are \cite{pdg}
\beq
m_{\nu_e}\lesssim 10\,{\mbox{eV}}, \qquad
m_{\nu_\mu}\leq 0.17\,{\mbox{MeV}}, \qquad
m_{\nu_\tau}\leq 18\,{\mbox{MeV}},
\eeq
and cosmological constraints
require stable neutrinos to be lighter than about $100\,$eV \cite{HaNi}.
However, the solar \cite{hata} and atmospheric \cite{atmo} 
neutrino puzzles, 
the LSND results \cite{LSND} and
models of mixed 
(hot and cold) dark matter \cite{Primack}
suggest that neutrinos may have
a small mass, $m_\nu \sim 10^{-5} - 10\,$eV.
The most attractive way to get such masses is to 
introduce Majorana neutrino mass terms that violate lepton number 
($L$) by two units.

In this talk, based upon Ref. \cite{gh,ghn}
(see also \cite{HKK} for an independent study), we consider
a supersymmetric Standard Model with
Majorana neutrino masses.
In such models, lepton number violation can
generate interesting phenomena in the slepton sector.
In additional to generating small neutrino masses,
the $\Delta L=2$ operators introduce a mass 
splitting and mixing in the sneutrino--antisneutrino system.
The sneutrino and antisneutrino will then no
longer be mass eigenstates.

This is  analogous to the neutral meson systems\cite{revB}.
For example, 
in the $B^0$ system the effect of a small $\Delta B=2$ perturbation
to the leading $\Delta B=0$ mass term
results in a mass splitting between the
heavy and light $B^0$, which are no longer pure $B^0$ and $\bar{B^0}$
states.  The very small mass splitting, ${\Delta m_B /m_B} = 6 \times
10^{-14}$ \cite{pdg}, can be measured by observing flavor oscillations.
The flavor is tagged in $B$-decays
by the final state lepton charge.
Since $x_d \equiv \Delta m_B /\Gamma_B \approx 0.7$ \cite{pdg}, 
there is time for
the flavor to oscillate before the meson decays.  
When $B$ mesons are produced in pairs
(for example in $e^+e^-$ collider operating at the $\Upsilon(4S)$ resonance)
the same sign dilepton signal indicates that only one of the $B$  
oscillated.
This time-integrated same
sign dilepton sample is used to determine the tiny mass splitting.

The sneutrino system can exhibit similar behavior. The lepton number is tagged
in sneutrino decay using the charge of the outgoing lepton.  
If the sneutrino has time to mix before it decays, namely if
\beq \label{xsnudef}
x_{\snu} \equiv {\dmsnu \over \Gamma_{\snu}} \gtrsim 1,
\eeq
and if the branching ratio of the sneutrino decay
into a charged lepton is significant,
then we can directly measure a non-zero sneutrino mass spliting via
the same sign dilepton signal.
When the sneutrinos are pair produced, {\it e.g.} in $e^+e^-$ collisions,
the two leptons from the sneutrino decays are used.
When the sneutrino is produced together with a charged lepton, 
{\it e.g.} in hadron collider via cascade decays,
the lepton from the sneutrino decay and the associated produced lepton
are used.
In both cases a measurable same sign dilepton signal is expected.

The neutrino mass and the
sneutrino mass splitting are both consequences of the small breaking
of lepton number. Therefore, they are expected to be
related. Thus, we can use upper bounds (or indications)
of neutrino masses to set bounds on the sneutrino mass splitting.
We will consider the consequences of two cases: (i) $\nu_\tau$ with a mass near 
its present laboratory upper limit, $m_\nu \sim 10 \,$MeV; 
and (ii) light neutrinos of mass less than $100\,$eV.

In order to derive specific results, one must specify a model for the
lepton number violation.  In the following we 
concentrate on two models
of neutrino masses: the see-saw mechanism and R-parity violation.
We compute the sneutrino mass splitting in each model
and its relation to the neutrino mass. We then 
briefly 
discuss the consequences for sneutrino phenomenology in colliders


\section{The Supersymmetric See-Saw Model}
Consider an extension of the MSSM where
one adds a right-handed neutrino
superfield, $\hat N$, with a bare mass $M \gg m_Z$.
We consider a one generation model
({\it i.e.}, we ignore lepton flavor mixing) and assume CP conservation.  We
employ the most general R-parity conserving
renormalizable superpotential and attendant
soft-supersymmetry breaking terms.  For this work, the relevant terms
in the superpotential are (following the notation of
Ref.~\cite{habertasi})
\beq
W=\epsilon_{ij}\left[\lambda \hat H_2^i \hat L^j \hat N - \mu \hat H_1^i
\hat H_2^j\right]+ \half M\hat N \hat N\,.
\eeq
The $D$-terms are the same as in the MSSM.
The relevant terms in the soft-supersymmetry-breaking scalar potential
are:
\beq
V_{{\rm soft}}= m_{\tilde L}^2 \snu^*\snu + m_{\tilde N}^2 \tilde N^*
\tilde N  + 
(\lambda A_\nu H_2^2\snu\tilde N^* 
+ MB_N \tilde N\tilde N + {\rm h.c.})\,.
\eeq
When the neutral Higgs field vacuum
expectation values are generated 
[$\langle H_i^i\rangle=v_i/\sqrt{2}$, with
$\tan\beta\equiv v_2/v_1$ and $v_1^2+v_2^2=v^2=(246~{\rm GeV})^2$], 
one finds that
the light neutrino mass is given by the usual one generation see-saw result
\beq \label{treemass}
m_\nu = {m_D^2 \over M}\,,
\eeq
where $m_D\equiv\lambda v_2$ and we drop terms higher order in $m_D/M$.

The sneutrino masses are obtained by diagonalizing a $4\times 4$ 
squared-mass matrix.  Here, it is convenient to define: 
$\snu=(\snu_1+i\snu_2)/\sqrt{2}$ and
$\tilde N=(\tilde N_1+i\tilde N_2)/\sqrt{2}$.  Then,
the sneutrino squared-mass matrix
separates into CP-even and CP-odd blocks:
\beq
-{\cal L}_{{\rm mass}}
=\half \pmatrix{\phi_1&\phi_2\cr}
\pmatrix {{\cal M}^2_+ & 0 \cr 0 & {\cal M}^2_-
\cr}\pmatrix{\phi_1\cr
\phi_2 \cr}\,,
\eeq
where $\phi_i\equiv \pmatrix{\snu_i&\tilde N_i\cr}$ and 
\beq
{\cal M}^2_\pm=\pmatrix{
m_{\tilde L}^2 \! + \! \half m_Z^2 \cos 2\beta\!  +\!  m_D^2 &
m_D[A_\nu-\mu \cot\beta\pm M] \cr
m_D[A_\nu-\mu \cot\beta\pm M] & M^2 \! +\!  m_D^2 \! 
+ \! m_{\tilde N}^2 \pm 2 B_N M  
\cr}\!.
\eeq
In the following derivation we assume that $M$ is the largest
mass parameter. Then, to first order in $1/M$,
the two light sneutrino
eigenstates are $\snu_1$ and $\snu_2$, with corresponding squared masses:
\beq
m^2_{\snu_{1,2}} = m_{\tilde L}^2+
\half m_Z^2 \cos 2\beta\mp\half\Delta m^2_{\snu}\,,
\eeq
where the squared mass difference $\Delta m^2_{\snu}\equiv
m^2_{\snu_2}-m^2_{\snu_1}$ is of order $1/M$. Thus, in the large $M$ limit, we
recover the two degenerate sneutrino states of the MSSM,
usually chosen to be $\snu$ and $\bar{\snu}$.  
For finite $M$, these two
states mix with a $45^\circ$ mixing angle, since the two
light sneutrino mass eigenstates
must also be eigenstates of CP.  The sneutrino mass splitting is easily
computed using $\Delta m^2_{\snu}=2m_{\snu}\Delta m_{\snu}$, where
$m_{\snu}\equiv\half(m_{\snu_1}+m_{\snu_2})$ is the average of the light
sneutrino masses. We find that the ratio of the light
sneutrino mass difference relative to the light {\it neutrino} mass 
[eq.~(\ref{treemass})] 
is given by (to leading order in $1/M$)
\beq \label{ratio}
r_\nu \equiv {\Delta m_{\snu} \over m_\nu} \simeq
{2 (A_\nu-\mu \cot\beta-B_N) \over m_{\snu}}\,.
\eeq

The magnitude of $r_\nu$ depends on various supersymmetric
parameters.
Naturalness constrains supersymmetric mass parameters
associated with particles with non-trivial electroweak quantum numbers
to be roughly of order $m_Z$ \cite{natural}.
Thus, we assume that
$\mu$, $A_\nu$, and $m_{\tilde L}$ are all of order the electroweak
scale.  The parameters $M$, $m_{\tilde N}$,
and $B_N$ are fundamentally different since
they are associated with the SU(2)$\times$U(1) singlet superfield
$\hat N$.  In particular, $M\gg m_Z$, since this drives the see-saw
mechanism.  Since $M$ is a supersymmetry-conserving parameter,
the see-saw hierarchy is technically natural.
The parameters $m_{\tilde N}$ and $B_N$ are
soft-supersymmetry-breaking parameters; 
their order of magnitude is less clear.
Since $\hat
N$ is an electroweak gauge group singlet superfield,
supersymmetry-breaking terms associated with it need
not be directly tied to the scale of electroweak symmetry breaking.
Thus, it is possible that $m_{\tilde N}$ and $B_N$ are much larger than $m_Z$.
Since $B_N$ enters directly into the formula for the
light sneutrino mass splitting [eq.~(\ref{ratio})],
its value is critical for sneutrino phenomenology.
If $B_N\sim{\cal O}(m_Z)$, then $r_\nu\sim {\cal O}(1)$, 
which implies that the sneutrino mass splitting is of order the {\it
neutrino} mass.  However, if $B_N\gg m_Z$, 
then the sneutrino mass splitting is
significantly enhanced.

\section{R-parity violation}
Consider an extension of the MSSM where R-parity is not imposed
(in this case, we do not add right handed neutrinos).
Again, we ignore lepton flavor mixing 
and assume CP conservation. 
In this model one neutrino mass arises at tree level 
from neutrino mixing with 
the neutralinos via  sneutrino VEVs or 
quadratic terms (``$\mu$-terms") in the superpotential \cite{bgnn}.
The sneutrino splitting arises from sneutrino mixing
with the Higgs fields \cite{ghn}.
The other neutrino masses and sneutrino splittings
arise at one loop \cite{bgnn2} and are not considered here.

In models without R-parity
there is a priori nothing to distinguish
the lepton-doublet supermultiplets $\hat L_i$ from the down-Higgs
supermultiplet $\hat H_d$, as both transform as $(2)_{-1/2}$ under
$SU(2)_L\times U(1)_Y$. With one generation there are  two $Y=-1/2$ doublets
which we denote by $\hat H_1$ and $\hat H_3$.
Then, in the superpotential
the single $\mu$-term of the
MSSM is now extended to a vector
\beq
W=-\epsilon_{ij} \mu_\alpha \hat H_\alpha^i \hat H_2^j,
\eeq
where $\alpha=1,3$ here and it what follows.
The trilinear terms in the superpotential
are irrelevant here. The $D$-terms are the same as in the MSSM.
The single SUSY breaking $B$ term of the MSSM
is also extended to a vector,
and SUSY breaking scalar masses are extended into a matrix
\beq
V_{{\rm soft}}=
m_{\alpha\beta}^2 H_\alpha H_\beta^* + m_2^2 |H_2|^2 +
\left(B_\alpha H_\alpha H_2 +{\rm h.c.}\right).
\eeq
where $H_1, H_2$ and $H_3$ are the 
neutral components of the scalar fields.
Finally, the single down type Higgs vev, $v_1$,
is also extended to a vector, $v_\alpha$.

To get the neutrino mass we consider the neutralino mass matrix \cite{bgnn}.
For simplicity we consider only one generation.
The full $5\times5$ tree-level 
neutralino mass matrix with rows and columns
corresponding to
$\{\tilde B,\tilde W_3,\tilde H_2^0,\tilde H_1^0, \tilde H_3^0\}$ is
\beq
M^{\rm (n)}=\pmatrix{
M_1&0&m_Zs{v_2 \over v}&-m_Zs{v_1 \over v}&-m_Zs{v_3 \over v}\cr
0&M_2&-m_Zc{v_2 \over v}&m_Zc{v_1 \over v}&m_Zc{v_3 \over v}\cr
m_Zs{v_2 \over v}&-m_Zc{v_2 \over v}&0&\mu_1&\mu_3\cr
-m_Zs{v_1 \over v}&m_Zc{v_1 \over v}&\mu_1&0&0\cr
-m_Zs{v_3 \over v}&m_Zc{v_3 \over v}&\mu_3&0&0\cr},
\eeq
where $c=\cos\theta_W$, $s=\sin\theta_W$ and $v^2=v_1^2+v_2^2+v_3^2$.
We define
\beq
\mu\equiv \left(\sum_\alpha\mu_\alpha^2\right)^{1/2},\qquad
v_d\equiv \left(\sum_\alpha v_\alpha^2\right)^{1/2},\qquad
\cos\xi\equiv {\sum_\alpha v_\alpha\mu_\alpha\over v_d\mu}.
\eeq
Note that $\xi$ measures the alignment of $v_\alpha$ and $\mu_\alpha$.
The product of the masses is then
\beq
\det M^{\rm n} = 
\left(m_Z^2 \mu^2 m_{\tilde \gamma}\right) \cos^2\beta \sin^2\xi,
\eeq
where $\tan\beta={v_2/v_d}$ and
$m_{\tilde \gamma} = \cos^2\theta_W M_1 + \sin^2\theta_W M_2$.
In the MSSM with R-parity \cite{habertasi}, where the neutrino would be
massless, the product of the four non-vanishing masses is
\beq
\det M^{\rm (n)}_0 = 
\mu\left(-M_1 M_2 \mu + \sin 2\beta m_Z^2 m_{\tilde \gamma}\right).
\eeq
To first order in the neutrino mass, the neutralino masses are unchanged
by the R-parity violating terms. Thus, we get \cite{enrico}
\beq
m_\nu = {\det M^{\rm (n)} \over \det M^{\rm (n)}_0} =
\rho_\nu m_Z \cos^2\beta \sin^2\xi,
\eeq
with
\beq
\rho_\nu =
{m_Z \mu m_{\tilde \gamma} 
\over
\left(-M_1 M_2 \mu + \sin 2\beta m_Z^2 m_{\tilde \gamma}\right)}.
\eeq
We find $\rho_\nu\sim 1$ for $\mu \sim M_1 \sim M_2 \sim m_Z$.

The sneutrino mass splitting is a result of the difference
in sneutrino mixing with the CP even and CP odd Higgs fields.
The mass-squared matrices are given by
\beqa
&&\!\!\!\!\!\! M^{\rm odd}=
{1 \over v^2} \pmatrix{
a_{12}^2 v_2^2 + a_{13}^2 v_3^2& a_{12}^2 v_1 v_2 &
         - a_{13}^2v_1 v_3\cr
a_{12}^2 v_1 v_2 & a_{12}^2 v_1^2 +  a_{23}^2 v_3^2&
         a_{23}^2 v_2 v_3\cr
 - a_{13}^2 v_1 v_3 &  a_{23}^2 v_2 v_3 & 
         a_{13}^2 v_1^2 + a_{23}^2 v_2^2 \cr}, \\
&&\!\!\!\!\!\! M^{\rm even}= {1 \over v^2} \times \nonumber \\
&&\!\!\!\!\!\! \pmatrix{
m_Z^2 v_1^2 \! + \!
a_{12}^2 v_2^2 \! + \! a_{13}^2 v_3^2& - (m_Z^2 \! + \! a_{12}^2) v_1 v_2 &
         - (m_Z^2 \! + \!a_{13}^2) v_1 v_3\cr
- (m_Z^2 \! + \! a_{12}^2)v_1 v_2 & a_{12}^2 v_1^2 \! + \! m_Z^2 v_2^2 
\! + \! a_{23}^2 v_3^2&
         - (m_Z^2 \! + \! a_{23}^2)v_2 v_3\cr
 - (m_Z^2 \! + \! a_{13}^2)v_1 v_3 &  - (m_Z^2 \! + \! a_{23}^2)v_2 v_3 & 
         a_{13}^2 v_1^2 \! + \! a_{23}^2 v_2^2 \! + \! m_Z^2 v_3^2\cr}
\nonumber
\eeqa
where
\beq
a_{12}^2 \equiv {B_1 v^2 \over v_1 v_2}, \qquad
a_{23}^2 \equiv {B_3 v^2 \over v_2 v_3}, \qquad
a_{13}^2 \equiv {m_{13}^2 v^2 \over v_1 v_2}.
\eeq
Note that $M^{\rm odd}$
includes the massless Goldstone boson and two massive CP-odd scalares.
We work in the basis where $\mu_3=0$. Then,
$v_3 \to 0$ only when both $m_{13}^2  \to 0$ and $B_3 \to 0$ in such a way
that $a_{ij}$ is finite \cite{bgnn}. Thus, the only
small parameter is $v_3$.
We use $v_3 = v \cos\beta \sin\xi$ and find that 
to lowest order in $\sin\xi$
\beq
\Delta \msnu = \rho_{\snu}\, m_Z \cos^2\beta \sin^2\xi, 
\eeq
where $\rho_{\snu} \sim \rho_\nu$ is given in \cite{ghn}.
%
In particular, we find
\beq
r_\nu \equiv {\Delta m_{\snu} \over m_\nu} = 
{\rho_{\snu} \over \rho_{\nu}} \sim 1.
\eeq
Thus, we conclude that in models where
R-parity violation is the only source of lepton number violation,
$r_\nu\simeq{\cal O}(1)$, and no enhancement of the sneutrino mass
splitting is possible.


\section{Loop Effects}
In the previous sections, we took into account only tree level
contributions to the neutrino and sneutrino mass matrices.
However, in some cases, one-loop effects can substantially modify
$r_\nu$.
In general, the existence of a sneutrino mass splitting
generates a one-loop contribution to the neutrino mass.
Note that this effect is generic, and is independent of the mechanism that
generates the sneutrino mass splitting.
Similarly, the existence of a Majorana
neutrino mass generates a one-loop contribution
to the sneutrino mass splitting. 
In models discussed in this paper we found that 
$r_\nu \gtrsim 1$ at tree level,
and 
therefore, the latter effect can be safely neglected. 
In contrast, the one-loop correction to the neutrino mass is potentially
significant, and may dominate the tree-level mass. 
We have computed exactly the one-loop contribution to the neutrino mass
[$m_\nu^{(1)}$] from neutralino/sneutrino loops shown 
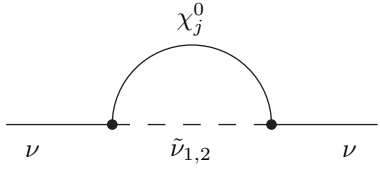
\begin{figure}
\begin{center}
\begin{picture}(200,80)(0,0)
\CArc(90,20)(30,0,180)
\DashLine(60,20)(120,20){7}
\Line(60,20)(20,20)
\Line(120,20)(160,20)
\Vertex(60,20){2}
\Vertex(120,20){2}
\Text(30,10)[]{$\nu$}
\Text(150,10)[]{$\nu$}
\Text(90,60)[]{$\chi^0_j$}
\Text(90,10)[]{$\tilde \nu_{1,2}$}
\end{picture}
\end{center}
\caption[a]{One-loop contribution to the neutrino mass due to
sneutrino mass splitting.}
\end{figure}
\noindent in Fig.~1. In the limit of
$m_\nu,\Delta m_{\snu}\ll m_{\snu}$,
the formulae simplify, and we find
\beq \label{loopmass}
m_\nu^{(1)} = {g^2\Delta m_{\snu} \over 32 \pi^2 \cos^2 \theta_W}
\sum_j \,f(y_j) |Z_{jZ}|^2\,,
\eeq
where
$f(y_j) =  \sqrt{y_j}\left[y_j-1-\ln(y_j)\right]/(1-y_j)^2$, with
$y_j \equiv {m_{\snu}^2/m_{\tilde\chi^0_j}^2}$, and
$Z_{jZ}\equiv Z_{j2}\cos\theta_W-Z_{j1}\sin\theta_W$ is the
neutralino mixing
matrix element that projects out the $\tilde Z$ eigenstate from the $j$th
neutralino. One can check that $f(y_j)<0.566$, and
for typical values of $y_j$ between 0.1 and 10, $f(y_j) > 0.25$.
Since $Z$ is a unitary
matrix, we find $m_\nu^{(1)}\approx  10^{-3} m_\nu^{(0)} r_\nu^{(0)}$, where
$r_\nu^{(0)}$ is the tree-level ratio. If
$r_\nu^{(0)}\gtrsim 10^3$, then the one-loop contribution to the neutrino mass
cannot be neglected.  Moreover, $r_\nu$ cannot be arbitrarily large
without unnatural fine-tuning.  Writing the neutrino mass as
$m_\nu= m_\nu^{(0)}+m_\nu^{(1)}$, and assuming no unnatural
cancellation between the two terms, we conclude that
\beq
r_\nu \equiv {\Delta m_{\snu} \over m_\nu} \lesssim 2\times 10^{3}.
\eeq


\section{Phenomenological Consequences}
Based on the analysis presented above, we take
$1 \lesssim r_\nu \lesssim 10^3$.  If $r_\nu$ is near its maximum, and if there
exists a neutrino mass in the MeV range, then the corresponding sneutrino mass
difference is in the GeV range.  Such a large mass splitting can be observed
directly in the laboratory.  For example, in $e^+e^-$ annihilation,
third generation sneutrinos
are produced via $Z$-exchange. Since the two sneutrino mass
eigenstates are CP-even and CP-odd respectively, 
sneutrino pair production
occurs only via $e^+e^-\to \snu_1\snu_2$.
In particular, the pair production processes 
$e^+e^-\to\snu_i\snu_i$ (for $i=1,2$) are forbidden.
If the low-energy supersymmetric model incorporates some R-parity
violation, 
then sneutrinos can be produced as an s-channel resonance
in $e^+e^-$ collisions \cite{EFP,BGH}.
Then, for a sneutrino mass difference in the GeV range,
two sneutrino resonant peaks could be distinguished.

A smaller sneutrino mass splitting can be probed 
using the same sign dilepton
signal if $x_{\snu} \gtrsim 1$.  Here we must rely on sneutrino oscillations.
Assume that the sneutrino decays with significant branching ratio via
chargino exchange:  $\snu \to \ell^\pm + X$.  Since this decay
conserves lepton number, the lepton number
of the decaying sneutrino is tagged by the lepton charge.  Then in
$e^+e^-\to\snu_1\snu_2$, the probability of a same sign dilepton signal
is
\beq
P(\ell^+\ell^+) + P(\ell^-\ell^-)= \chi_{\snu}
\left[\BR(\snu \to \ell^\pm +X)\right]^2\,,
\eeq
where 
\beq
\chi_{\snu} \equiv x_{\snu}^2/[2(1+x_{\snu}^2)],
\eeq
is the integrated oscillation probability, which
arises in the same way as the corresponding quantity that appears in the
analysis of $B$ meson oscillations \cite{revB}.
At hadron collider, where the sneutrino are produced
mainly via $\chi_2^+ \to \snu \ell^+$
the probability of a same sign dilepton signal is
\beq
P(\ell^+\ell^+) + P(\ell^-\ell^-)= \chi_{\snu}
\left[\BR(\snu \to \ell^\pm +X)\right]\,.
\eeq
We have considered the constraints on the supersymmetric model imposed
by the requirements that 
$x_{\snu}\sim{\cal O}(1)$ and BR$(\snu\to\ell^\pm+X)\sim 0.5$.
We examined two cases depending on
whether the dominant $\snu$ decays involve two-body or three-body final states.

If the dominant sneutrino decay involves two-body final states, then we must
assume that $m_{\tilde\chi_1^0}< m_{\tilde\chi^+} <m_{\snu}$.
Then, the widths of the two leading sneutrino decay channels,
with the latter summed over both final state charges,
are given by \cite{BGH,gunhab}
\beqa \label{gammas}
\Gamma(\snu \to \tilde\chi_j^0 \nu)& =&
{g^2|Z_{jZ}|^2 m_{\snu}\over 32 \pi \cos^2 \theta_W}
B(m_{\tilde\chi_j^0}^2/m_{\snu}^2)\,, \nonumber \\
\Gamma(\snu \to \tilde\chi^\pm \ell^\mp) &=& 
{g^2|V_{11}|^2 m_{\snu}\over 8 \pi}
B(m_{\tilde\chi^+}^2/m_{\snu}^2)\,,
\eeqa
where $B(x)\equiv (1-x)^2$,
$V_{11}$ is one of the mixing matrix elements in the chargino sector,
and $Z_{jZ}$ is the neutralino mixing matrix element
defined below eq.~(\ref{loopmass}), and we take $m_\ell=0$. For example,
for $m_{\snu} \sim {\cal O}(m_Z)$ we find
\beqa
\Gamma(\snu \to \chi_j^0 \nu) &\approx& {\cal O}\left(|Z_{jZ}|^2
B(m_{\tilde\chi_j^0}^2/m_{\snu}^2)\times 1\,
{\mbox{\rm GeV}}\right) \\
\Gamma(\snu \to \chi^+ \ell) &\approx&  {\cal O}\left(|V_{11}|^2
B(m_{\tilde\chi^+}^2/m_{\snu}^2)\times 1 \,{\mbox{\rm GeV}}\right) \nonumber.
\eeqa
Typically, $B\gtrsim 10^{-2}$ in eq.~(\ref{gammas}). Thus,
for the third generation sneutrino, a significant same-sign dilepton
signal can be generated with $m_{\nu_\tau}= 10\,$MeV,
even if $r_\nu \sim 1$
and the light chargino/neutralino mixing angles are of ${\cal O}(1)$.
If the lightest chargino and two lightest neutralinos
are Higgsino-like, then the mixing angle factors in eq.~(\ref{gammas}) are
suppressed.  For $|\mu|\sim m_Z$ and gaugino mass parameters not larger
than $1\,$TeV, the square of the light chargino/neutralino mixing angles
must be of ${\cal O}(10^{-2})$ or larger.  Thus, if
$r_\nu$ is near its maximum value ($r_\nu \sim 10^3$), then one
can achieve $x_{\snu} \sim 1$ for {\it neutrino} masses as low as about
$100\,$eV.

If no open two-body decay channel exists, then we must consider the
possible sneutrino decays into three-body final states.  In this case
we require that $m_{\snu}<m_{\tilde\chi_1^0},m_{\tilde\chi^+}$.  Again, we
assume that there exists a significant chargino-mediated
decay rate with charged leptons in the final state.
The latter occurs in models in which the $\tilde\tau_R$
is lighter than the sneutrino.  In this case, the rate for chargino-mediated
three-body decay $\snu_\ell\to\tilde\tau_R\nu_\tau\ell$ can be
significant.  The $\tilde\tau_R$ with $m_{\tilde\tau_R}<m_{\snu}$
can occur in
radiative electroweak breaking models of low-energy supersymmetry if
$\tan\beta$ is large.  However, in the context of the MSSM, such a
scenario would require that $\tilde\tau_R$ is the lightest
supersymmetric particle (LSP), a possibility strongly disfavored
by astrophysical bounds on the abundance of stable heavy charged particles.
Thus, we go beyond the usual MSSM assumptions and assume
that the $\tilde\tau_R$ decays.  This can occur in gauge-mediated
supersymmetry breaking models \cite{DDRT} where
$\tilde\tau_R\to\tau \tilde g_{3/2}$, or in R-parity violating
models where $\tilde\tau_R \to\tau\nu$.
Here, we have assumed that intergenerational lepton
mixing is small;
otherwise the $\Delta L=2$ sneutrino mixing effect is diluted.

We have computed the chargino and neutralino-mediated three-body
decays of $\snu_\ell$.
In the analysis presented here, we have not considered the case of
$\ell=\tau$, which involves a more complex final
state decay chain containing two $\tau$-leptons.
For simplicity, we present analytic formulae in the limit where the
mediating chargino and neutralinos are much heavier than the
$\tilde\tau_R$.  In addition, we assume that the
lightest neutralino is dominated by its bino component.
We have checked that our conclusions
do not depend strongly on these approximations.
Then, the rates for the chargino and neutralino-mediated
sneutrino decays 
(the latter summed over both final state charges) are
\beqa
\Gamma(\snu_\ell \to \ell^-\tilde\tau^+ \nu_\tau) &=&
{g^4 m_{\snu}^3 m_\tau^2 \tan^2\beta\,
f_{\tilde \chi^+}(m_{\tilde \tau}^2/m_{\snu}^2) \over
1536  \pi^3 (m_W^2 \sin 2\beta - M_2 \mu)^2}\,, \nonumber \\
\Gamma(\snu_\ell \to \tau^\pm \tilde \tau^\mp\nu_\ell)&=&
{g'^4 m_{\snu}^5\, f_{\tilde
\chi^0}(m_{\tilde \tau}^2/m_{\snu}^2)  \over
3072  \pi^3 M_1^4} \,,
\eeqa
for $\ell=\mu$, $e$,
where the $M_i$ are gaugino mass parameters and
\beqa
f_{\tilde \chi^+}(x) &=& (1-x)(1+10 x+x^2) + 6x(1+x) \ln x, \\
f_{\tilde \chi^0}(x) &=& 1-8 x + 8 x^3 - x^4 - 12 x^2 \ln x. \nonumber
\eeqa
As an example, for $\tan\beta=20$ (consistent with a light
$\tilde\tau_R$ as noted above) and $m_{\tilde \tau}^2/m_{\snu}^2=0.64$, 
reasonable values for the
other supersymmetric parameters can be found such that
$\Gamma(\snu_\ell \to \ell^\pm\tilde \tau^\mp \nu_\tau) \sim
\Gamma(\snu_\ell \to \tau^\pm\tilde\tau^\mp\nu_\ell)\sim {\cal O}
(1\,{\mbox{\rm eV}})$.  In this case,
for $r_\nu \sim 1$ [$10^3$], a significant like-sign dilepton signal
could be observed for light neutrino masses as low as $1\,$eV [$10^{-3}\,$eV].


\section{Conclusions}
Non-zero Majorana neutrino masses imply the existence of
$\Delta L=2$ phenomena. 
In particular, in supersymmetric models, we expect
sneutrino-antisneutrino mixing. The resulting sneutrino mass
splitting is generally of the same order as the light
neutrino mass, although an enhancement of up to three
orders of magnitude is conceivable.
If the mass of the tau neutrino is near its present experimental bound,
$m_\nu \sim 10\,$MeV,
then it may be possible to directly observe the sneutrino mass splitting
in the laboratory.  Even if neutrino masses are small (of order $1\,$eV),
some supersymmetric models yield an observable sneutrino oscillation
signal.  Remarkably, model parameters exists
where sneutrino mixing phenomena are detectable for {\it neutrino}
masses as low as $m_\nu \sim 10^{-3}\,$eV
(a mass suggested by the solar neutrino
anomaly).  Thus, sneutrino mixing and oscillations could provide a novel
opportunity to probe lepton-number violating phenomena in the laboratory.

\section*{Acknowledgments}
I thank  Howie Haber and Yossi Nir for collaboration on this work and
Jens Erler, Jonathan Feng, Enrico Nardi, Scott Thomas and Jim Wells
for helpful discussions.  
Y.G. is supported by the Department of Energy under contract
DE-AC03-76SF00515. 
%



\end{document}